\documentclass[11pt,twoside]{article}
\usepackage{asp2006}
\usepackage{psfig}
\usepackage{epsf}
\usepackage{graphics}
\usepackage{lscape}
\def\edcomment#1{\iffalse\marginpar{\raggedright\sl#1\/}\else\relax\fi}
\reversemarginpar

\begin{document}
\title{Not only once: the amazing $\gamma$-ray activity of the blazar PKS 1510$-$089}
\author{F. D'Ammando\altaffilmark{1,2}, C. M. Raiteri\altaffilmark{3},
  M. Villata\altaffilmark{3}, P. Romano\altaffilmark{4},
  S. Covino\altaffilmark{5}, H. Krimm\altaffilmark{6},
  M. Tavani\altaffilmark{1,2}, and the AGILE and GASP-WEBT Collaborations}
\altaffiltext{1}{Univ. degli Studi di Roma Tor Vergata, Via della Ricerca Scientifica 1,
  I-00133, Roma Italy}
\altaffiltext{2}{INAF-IASF Roma, Via Fosso del Cavaliere 100, I-00133 Roma, Italy}
\altaffiltext{3}{INAF, Oss. Astronomico di Torino, Via Osservatorio 20, I-10025 Pino
  Torinese (Torino), Italy }
\altaffiltext{4}{INAF-IASF Palermo, Via Ugo La Malfa 153, I-90146 Palermo, Italy}
\altaffiltext{5}{INAF, Oss. Astronomico di Brera, Via Bianchi 46, I-23807 Merate (LC),
  Italy}
\altaffiltext{6}{NASA/Goddard Space Flight Center, Greenbelt, MD 20771, USA}

\begin{abstract} PKS 1510$-$089 is a powerful Flat Spectrum Radio Quasar at
  z=0.361 with radiative output dominated by the $\gamma$-ray emission. In the
  last two years PKS 1510$-$089 showed high variability over all the
  electromagnetic spectrum and in particular very high $\gamma$-ray activity was
  detected by the Gamma-Ray Imaging Detector on board the AGILE satellite with
  flaring episodes in August 2007 and March 2008. An extraordinary activity
  was detected in March 2009 with several flaring episodes and a flux that reached
  500 $\times$ 10$^{-8}$ photons cm$^{-2}$ s$^{-1}$. Multiwavelength observations of PKS
  1510$-$089 seem to indicate the presence of Seyfert-like features such as
  little blue bump and big blue bump. Moreover, X-ray observations suggested
  the presence of a soft X-ray excess that could be a feature of the bulk
  Comptonization mechanism. We present the results of the analysis of the
  multiwavelength data collected by GASP-WEBT, REM, {\it Swift} and AGILE during these $\gamma$-ray flares and the theoretical implications for the emission mechanisms.
\vspace{-0.5cm}

\end{abstract}


\section{Introduction}
Blazars are the most extreme class of Active Galactic Nuclei (AGNs) characterized by the emission of strong
non-thermal radiation across the entire electromagnetic spectrum, from radio
to very high $\gamma$-ray energies. The typical observational properties of
blazars include
irregular, rapid and often very large variability, apparent super-luminal
motion of the jet at VLBI scales, flat radio spectrum, high and variable polarization at radio and
optical frequencies. These features are interpreted as the result of
the emission of electromagnetic radiation from a relativistic jet that is
viewed closely aligned to the line of sight, thus causing strong relativistic amplification (Blandford $\&$ Rees 1978; Urry
$\&$ Padovani 1995).

The EGRET instrument onboard $Compton$ $Gamma$-$Ray$ $Observatory$ detected
for the first time strong and variable $\gamma$-ray
emission from blazars in the MeV-GeV region. In conjunction with ground-based
Cherenkov telescopes and coordinated multiwavelength observations, this provided
evidence that the Spectral Energy Distributions (SEDs) of blazars
are typically double-humped with the first peak occurring in the IR/optical band
for the so-called $red$ $blazars$ (including Flat Spectrum Radio Quasars, FSRQs, and
Low-energy peaked BL Lacs, LBLs) and at UV/X-rays for the so-called $blue$
$blazars$ (including High-energy peaked BL Lacs, HBLs). The first peak is
commonly interpreted as synchrotron radiation from high-energy electrons in
a relativistic jet. The SED second component, peaking at MeV-GeV energies in
$red$ $blazars$ and at TeV energies in $blue$ $blazars$, is commonly interpreted as
inverse Compton (IC) scattering of seed photons, internal or external to the jet, by relativistic electrons
(Ulrich et al., 1997), although other models have been
proposed (see e.g., B\"ottcher 2007 for a recent review). Blazars with greater
bolometric luminosity have smaller peak frequencies and ``redder'' SEDs, while
blazars of lower bolometric luminosity have higher peak frequencies and then
are ``bluer'' (Fossati et al., 1998). This spectral sequence was interpreted by
Ghisellini et al.~(1998)  as consequence of the different radiative cooling
suffered by the emitting electrons of blazars of different power. 

With the detection of several blazars in the $\gamma$-rays by EGRET (Hartman et
al., 1999) the study of this class of objects has made
significant progress. In fact, considering that the large fraction of the
total power of blazars is emitted in the $\gamma$-rays, information in
this band is crucial to
study the different radiation models. 
The interest in blazars is now even more renewed thanks to the simultaneous presence
of two $\gamma$-ray satellites, AGILE and $Fermi$, and the possibility to
obtain $\gamma$-ray
observations simultaneously with
multiwavelength data collected from radio to TeV energies will allow us to reach a deeper insight on the jet
structure and the emission mechanisms at work in blazars.

\section{PKS 1510$-$089}

PKS 1510$-$089 is a highly polarized radio-loud quasar at redshift z = 0.361 belonging to
the class of the FSRQs with radiative output dominated by the $\gamma$-ray emission, while the synchrotron emission peaks around IR
frequencies below a pronunced UV bump, likely due to the thermal
emission from the accretion disc (Malkan $\&$ Moore 1986; Pian $\&$ Treves
1993). Its radio emission exhibited very rapid, large amplitude variations in
both total and polarized flux (Aller, Aller, $\&$ Hughes 1996). The radio jet
shows superluminal motion up to 20$c$, with the parsec
and kiloparsec scale jets misaligned of 177 degrees (Wardle et al., 2005).

PKS 1510$-$089 has been extensively observed in X-rays in the last three decades
from the first observation by $Einstein$ (Canizares and White 1989) up to a long-term monitoring
performed by $Swift$ in July--August 2009 in the context of an extensive
multifrequency campaign on this source, whose results will be presented in a forthcoming paper
(D'Ammando et al., in preparation). 

The X-ray spectrum of the source observed by ASCA in the 2--10 keV band
(Singh, Shrader, $\&$ George 1997) was very flat (with photon index $\Gamma$ $\simeq$ 1.3), but steepened in the $ROSAT$ bandpass below 2 keV ($\Gamma$ $\simeq$
1.9), suggesting the possible presence of a spectral break around 1--2 keV,
associated with the existence of a soft X-ray excess. Observations by $BeppoSAX$
(Tavecchio et al., 2000) and Chandra (Gambill et al., 2003) confirms the presence
of a soft X-ray excess below 1 keV. A similar soft excess has been detected in
other blazars such as 3C 273, 3C 279 and 3C 454.3, even if the origin of the soft X-ray
excess is still an open issue, not only for blazars but for all AGNs (see
e.g. D'Ammando et al., 2008a for a discussion on the soft X-ray excess problem
in the radio-quiet AGNs). 

A monitoring campaign on this source was organized during August 2006 by
$Suzaku$ and $Swift$. The $Suzaku$ observations confirm the presence of a soft
X-ray excess, suggesting that it could be a feature of the bulk Comptonization, whereas the
$Swift$/XRT observations reveal significant spectral evolution of the X-ray
emission on timescales of a week: the X-ray spectrum becomes harder
as the source gets brighter (Kataoka et al., 2008).

Gamma-ray emission from PKS 1510$-$089 has been detected in the past by EGRET
during low/intermediate states with an integrated flux above 100 MeV varying between
(13 $\pm$ 5) and (49 $\pm$ 18) $\times$ 10$^{-8}$ photons cm$^{-2}$
s$^{-1}$. Instead in the last two years this source showed an intense $\gamma$-ray
activity with several flaring episodes detected by AGILE and $Fermi$.

In this paper we present the results of the analysis of the AGILE, {\it Swift}, GASP-WEBT and REM data obtained
during the multiwavelength observations of PKS 1510$-$089 performed in August
2007, March 2008, and March 2009. Throughout the paper the quoted
uncertainties are given at 1-$\sigma$ level, unless otherwise stated.

\section{August 2007}

PKS 1510$-$089 was detected for the first time by AGILE in high $\gamma$-ray activity during the period 23 August -- 1
September 2007, as reported in Pucella et al.~(2008). AGILE detected the source in a very bright state, with an average flux of
$F_{E>100 MeV}$ = (195 $\pm$ 30) $\times$ 10$^{-8}$ photons cm$^{-2}$ s$^{-1}$
between 28 August and 1 September 2007.

The simultaneous optical monitoring of the GLAST AGILE Support Program (GASP) of the Whole
Earth Blazar Telescope (WEBT) showed that PKS 1510$-$089 was in optical
decrease during the period of the AGILE observation,
following a bright state in mid August with the source at $R$ = 15.0 mag. The SED is
modelled with a synchrotron self Compton (SSC) +
external Compton (EC) emissions. The IC scattering of external photons from the
BLR seems to
explain the observed hard $\gamma$-ray spectrum observed by AGILE. 

After this flaring episode many more $\gamma$-ray flares of PKS 1510$-$089 were
detected by the AGILE and $Fermi$ satellites between March 2008 (D'Ammando
et al., 2008b) and April 2009 (Cutini et al., 2009).

\section{March 2008}

During a pointing toward the Galactic Center region, between 1 and 30 March
2008, AGILE detected a rapid $\gamma$-ray flare from PKS 1510$-$089, as
discussed in detail in D'Ammando et al.~(2009a). After two episodes of medium intensity the source
was not detected for some days in the $\gamma$-ray band and suddenly a rapid flare
was observed by AGILE on 18-19 March (see Fig. 1, left panel). 

During the period 1--16 March 2008, AGILE detected
an average flux from PKS 1510$-$089 of (84 $\pm$ 17) $\times$ 10$^{-8}$ photons
cm$^{-2}$ s$^{-1}$ for E $>$ 100 MeV. The flux measured between 17 and 21 March
was a factor of 2 higher, with a peak level of (281 $\pm$ 68) $\times$ 10$^{-8}$ photons
cm$^{-2}$ s$^{-1}$ on 19 March 2008.

Moreover, between January and April 2008 the source showed an
intense and variable optical activity with several flaring episodes of fast
variability. Peaks were detected by GASP-WEBT in the optical band on 15 February, 29 March and 11 April 2008 (see
Fig. 1, right panel). A significant increase of the flux was observed also at millimetric
frequencies in mid April, suggesting that the mechanisms producing the flaring events in
the optical and $\gamma$-ray bands also interested the millimetric zone, with a delay.

\begin{figure}[!h]
\plottwo{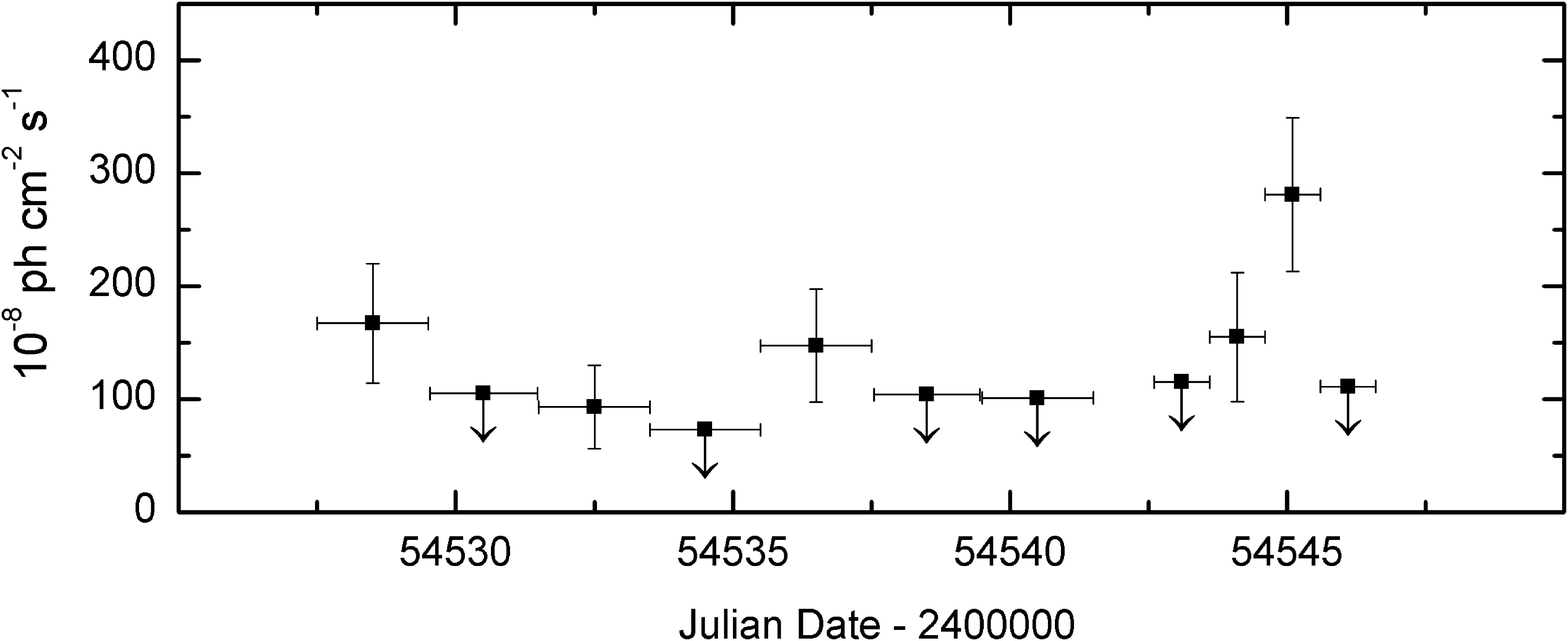}{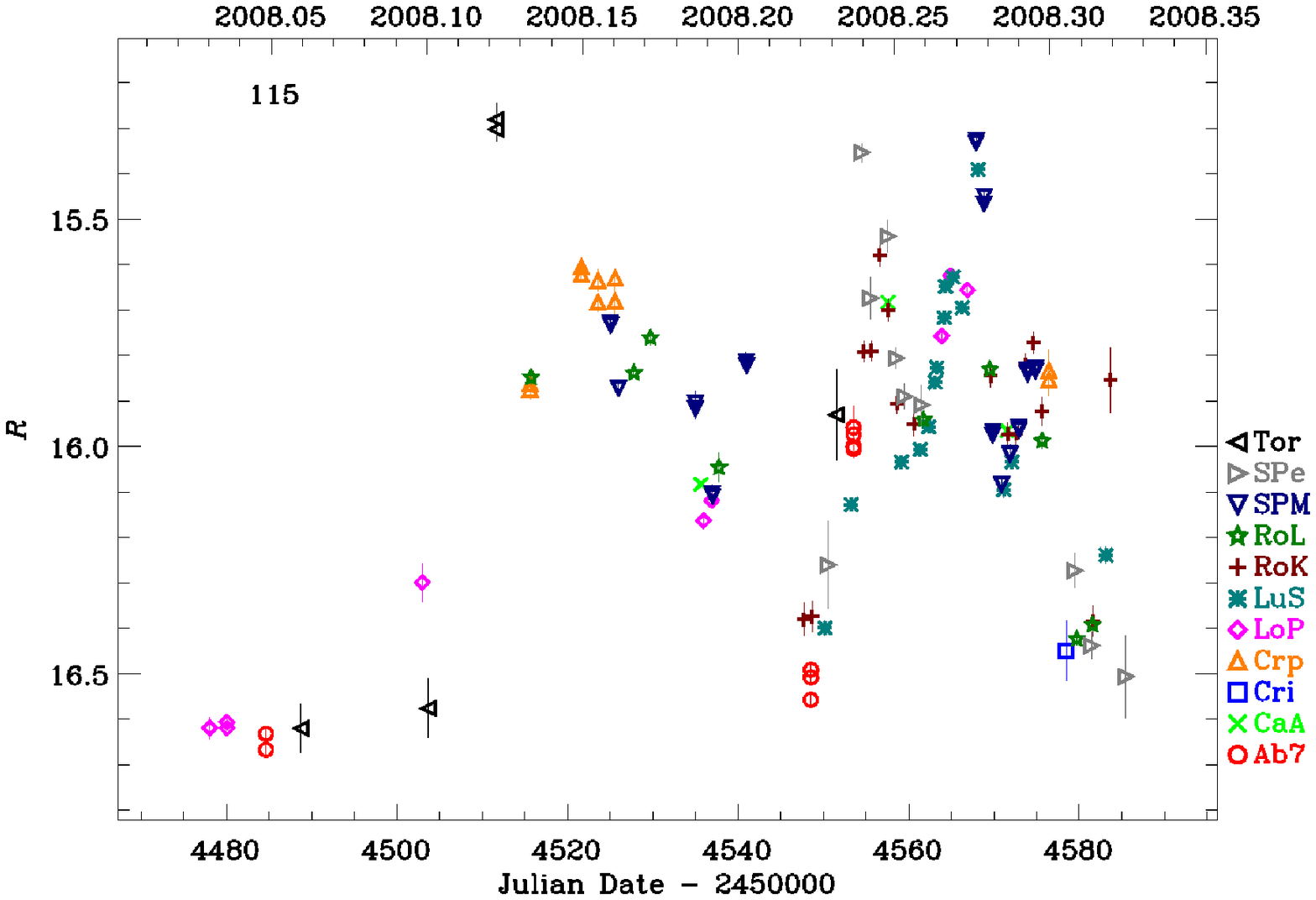}
\caption{{\it Left panel}: AGILE $\gamma$-ray light curve between 1 and 21 March
  2008 at 1 or 2-day resolution for E $>$ 100 MeV. The downward arrows
  represent 2-$\sigma$ upper limits. {\it Right panel}: $R$-band light curve
 obtained by GASP-WEBT during the period January-April 2008. Different symbols
  refer to different observatories.} 
\end{figure}

The $\gamma$-ray flare triggered 3 ToO observations with $Swift$ in three
consecutive days between 20 and 22 March 2008. The first XRT observation showed a very hard X-ray
photon index ($\Gamma$ = 1.16 $\pm$ 0.16) with a flux in the 0.3--10 keV band of (1.22 $\pm$ 0.17) x 10$^{-11}$ erg
cm$^{-2}$ s$^{-1}$ and a decrease of the flux of about 30$\%$ between 20 and
21 March. 
Moreover, the $Swift$/XRT observations seem to show a harder-when-brighter behaviour of the
spectrum in the X-ray band, confirming a behaviour already observed in this
source by Kataoka et al.~(2008). This is a trend usually observed in HBL such as Mrk 421
(see e.g. Tramacere et al., 2007) but quite rare in FSRQs such as PKS
1510$-$089. This harder-when-brighter behaviour is likely due to the different
variability of the SSC and EC components, therefore to the change of the
relative contribution of each component.
Thus, the X-ray photon index observed on 20 March could be due to the combination of SSC and EC emission and therefore to the mismatch of the
spectral slopes of these two components.

The SED for the AGILE observation of 17--21 March 2008 together with the
simultaneous data collected in radio, mm, near-IR, and optical bands by
GASP-WEBT and UV and X-ray bands by $Swift$ is modelled with thermal emission
of the disc, SSC model plus the contribution by the external Compton
scattering of direct disc radiation and of photons reprocessed by the Broad
Line Region (BLR; see Fig. 2, left panel). Some features in the optical-UV
spectrum seem to indicate the presence of Seyfert-like components, such as the
little and big blue bumps (see also Fig. 2, right panel). 

\begin{figure}[!th]
\plottwo{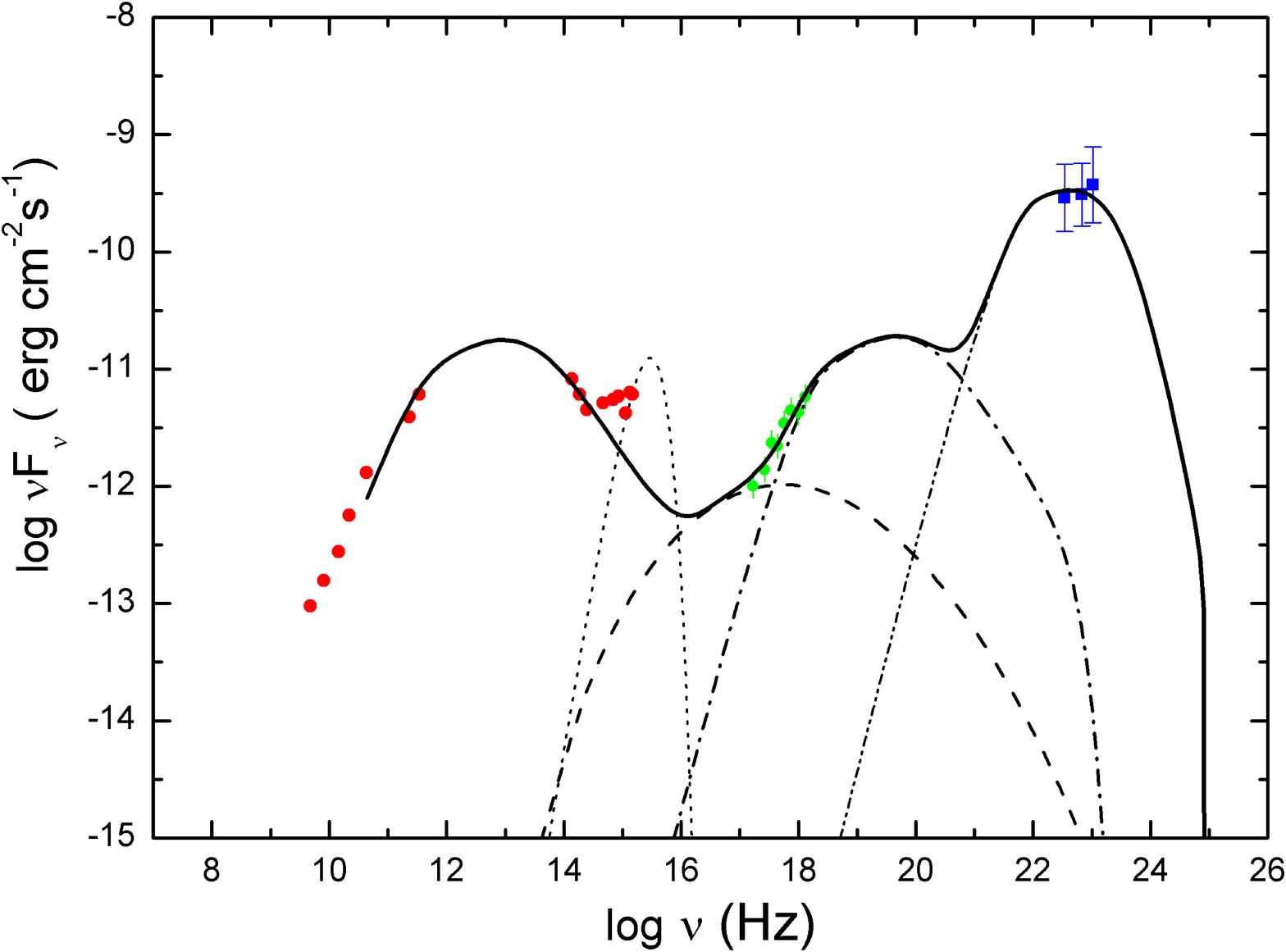}{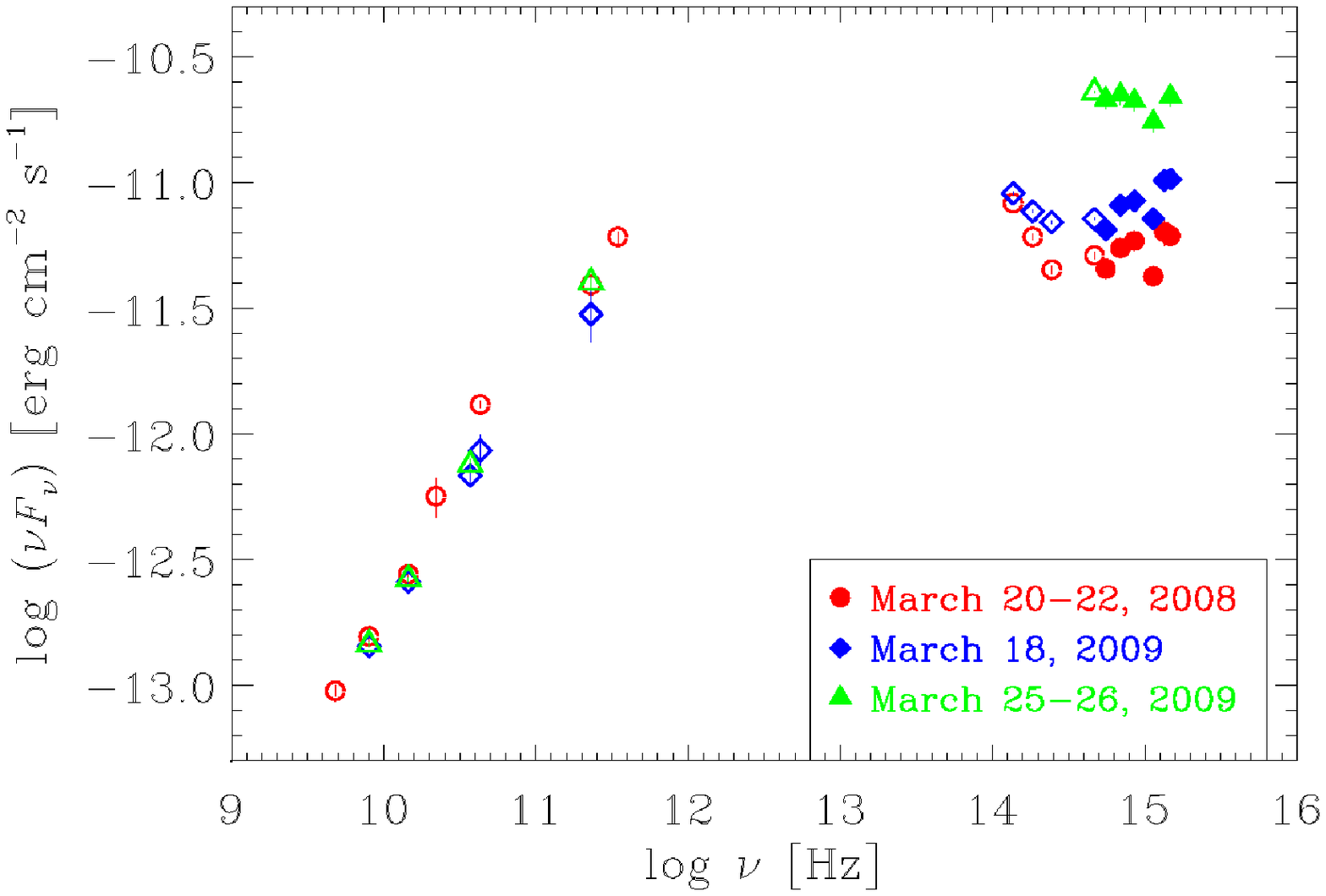}
\caption{{\it Left panel}: Broad-band SED of PKS 1510$-$089 for the AGILE observation of
  17--21 March 2008 including quasi-simultaneous radio-to-optical GASP data, UV
  and X-ray $Swift$ data. The dotted, dashed, dot-dashed and double-dot-dashed
  lines represent accretion disc black body, the SSC, the EC of the disc and
  of the BLR, respectively. {\it Right panel}: SED of the low-energy part of
  the spectrum constructed with data collected by GASP-WEBT and $Swift$/UVOT
  during March 2008 and March 2009.} 
\end{figure}

\section{March 2009}

PKS 1510$-$089 showed an extraordinary $\gamma$-ray activity during March
2009, with several flaring episodes that could be an overlapping of
different events (see Fig. 3, left panel). After a low intensity period in
February 2009, the optical activity of the source is also greatly increased in
March 2009 with an intense flare on 25--26 March (see Fig. 3 right panel). A similar behaviour
was observed by the REM Telescope, with an achromatic variation in the
near-infrared and optical bands. Taking into account that the dip at the UVW1
frequency it is also found for other blazars with different redshift and could
be systematic, the data collected from radio-to-UV on 25-26
March 2009 (Fig. 3, right panel) seem to show a flat spectrum in the
optical/UV energy band, suggesting an important contribution of the synchrotron emission
in this part of the spectrum during the huge flaring episode and therefore a
possible shift of the synchrotron peak, usually observed in this source in the
infrared band.

\begin{figure}[!t]
\plottwo{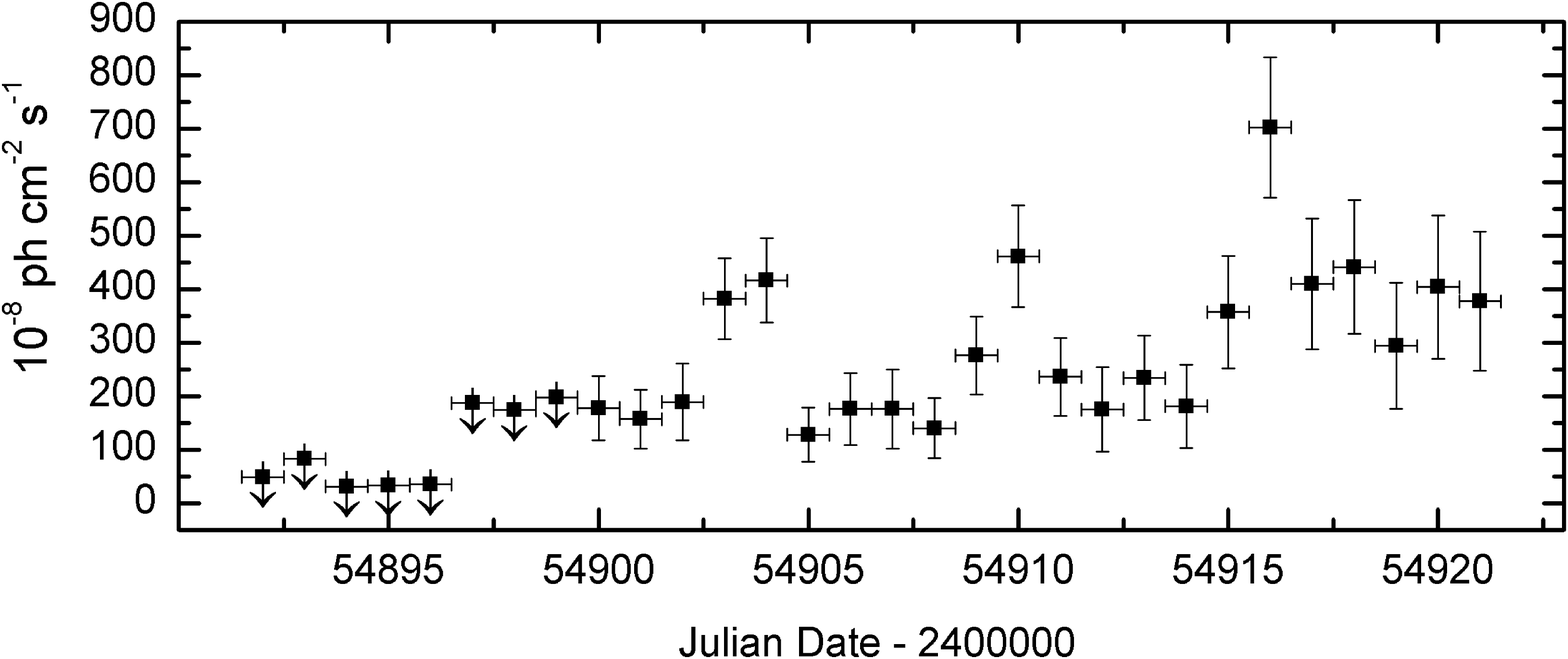}{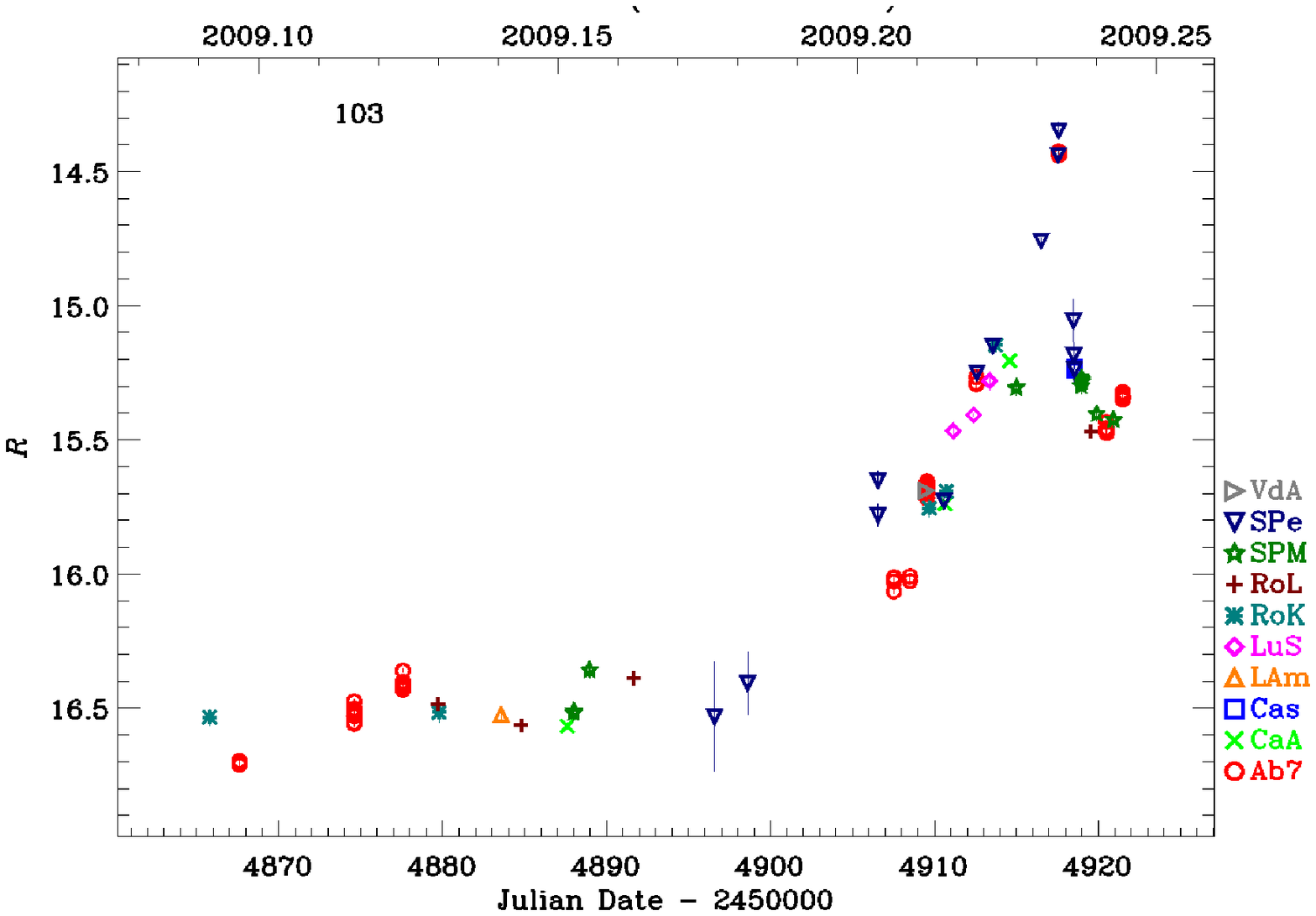}
\caption{{\it Left panel}: AGILE $\gamma$-ray light curve between 1 and 30 March
  2009 at 1-day resolution for E $>$ 100 MeV. The downward arrows
  represent 2-$\sigma$ upper limits. {\it Right panel}: $R$-band light curve
 obtained by GASP-WEBT during the period February-March 2009. Different symbols
  refer to different observatories.} 
\end{figure}

During the 14 ToO observations performed in March 2009, $Swift$/XRT
observed the source in an intermediate state with a 0.3--10 keV flux in the
range (7.5 -- 10.8) $\times$ 10$^{-12}$ erg cm$^{-2}$ s$^{-1}$. The X-ray flux
seems not to be correlated with the high optical and $\gamma$-ray activity. 
A hard X-ray outburst of this source was detected by $Swift$/BAT on 9 March
2009, with a rise from 15 mCrab to 40 mCrab in 24 hours. On 10 March the
source faded below the BAT sensitivity (Krimm et al., 2009). It is interesting to note that this
outburst in the 15--50 keV energy band occurred just at the beginning of the $\gamma$-ray activity observed by AGILE. 
The results of the multiwavelength campaign of March 2009 will be presented in D'Ammando
et al.~(2009b).

\acknowledgments AGILE is a mission of ASI, with co-participation of INAF and
INFN. This work is partly based on optical and near-infrared data provided by
the following GASP-WEBT observatories: Abastumani, Calar Alto, Campo Imperatore, Castelgrande, L'Ampolla,
Lulin, Roque de los Muchachos (KVA and Liverpool), San Pedro Martin, St.
Petersburg, Valle D'Aosta. This work is partly based on mm and radio data
provided by the following GASP-WEBT observatories: Medicina, Mets\"ahovi, Noto, SMA,
UMRAO. We thank the GASP-WEBT Collaboration for providing the data.

\end{document}